\documentclass[runningheads]{llncs}
\usepackage{wrapfig}
\usepackage{xcolor}
\usepackage{tabularx}
\usepackage{amsmath}
\usepackage{graphicx}
\usepackage{hyperref}
\usepackage{mathtools}
\usepackage{booktabs}
\usepackage{subcaption}
\usepackage{tcolorbox}
\usepackage{listings}

\usepackage[T1]{fontenc}
\lstdefinestyle{promptstyle}{
    backgroundcolor=\color{black!5},
    basicstyle=\ttfamily\scriptsize,
    breaklines=true,
    captionpos=b,
    frame=single,
    keepspaces=true,
    numbers=left,
    numbersep=5pt,
    rulecolor=\color{black!20},
    showstringspaces=false,
    tabsize=2,
    title=\lstname,
    literate=
      {á}{{\'a}}1 {é}{{\'e}}1 {í}{{\'i}}1 {ó}{{\'o}}1 {ú}{{\'u}}1
      {Á}{{\'A}}1 {É}{{\'E}}1 {Í}{{\'I}}1 {Ó}{{\'O}}1 {Ú}{{\'U}}1
      {ñ}{{\~n}}1 {Ñ}{{\~N}}1 {¿}{{?`}}1 {¡}{{!`}}1
}
\usepackage{lipsum} 

\usepackage{marvosym}
\begin{document}
\title{Conversational Retrieval and On-the-Fly Knowledge Modeling of Historical Penitentiary Repression Records}
\titlerunning{On-the-Fly Knowledge Modeling}
%
\author{Paula Font Solà, Adrià Molina Rodriguez\href{mailto:amolina@cvc.uab.cat}{\Letter}, Josep Lladós Canet}
\authorrunning{P. Font et.al}
%
\institute{Computer Vision Center, Universitat Autònoma de Barcelona\\\email{paula.fonts@autonoma.cat, \{amolina, josep\}@cvc.uab.cat}\\
}
\maketitle              
\begin{abstract}
Recent developments in digital libraries increasingly favor conversational and natural language access to information through Retrieval-Augmented Generation (RAG). Although these approaches are effective for extractive tasks grounded in individual records, they remain limited in their ability to interpret document collections holistically and to incorporate expert knowledge dynamically. In this article, we present a document analysis system designed for the management of historical digital libraries that supports on-the-fly knowledge modeling. The system is equipped with the capability to store facts produced either by expert archivists or derived from document retrieval processes within a graph-based structure. Through continuous professional interaction, the system can retrieve information not only from primary sources such as documents, but also from previously modeled knowledge, with the graph-based index acting as a memory for the language model to access. This enables increasingly complex queries involving long-term dependencies across documents, link discovery, and the integration of expert knowledge that may not be explicitly present in the original sources. As a result, the proposed approach facilitates the generation of richer and more comprehensive information.
\keywords{Document Analysis Systems \and Historical Documents}
\end{abstract}

\section{Introduction}
In recent years, historical document analysis systems have shifted from indices and enumerations of content to increasingly conversational and adaptive interfaces. This evolution is not merely a superficial redesign of how users access archives and portals; it represents a deeper transformation in how archival data is organized, interpreted, and communicated. The advent of Generative AI has accelerated this transition, transforming both how users interact with systems, using natural language queries and conversational exploration, and how results are synthesized and delivered. Instead of static document lists, modern systems generate summaries and reveal relationships in archival collections.  Although it may be tempting to attribute this transformation solely to the rapid development of Large Language Models (LLMs) and Retrieval-Augmented Generation (RAG), it is important to recognize that archival science had already begun moving beyond document enumeration toward more holistic paradigms. The International Council on Archives’s Records in Contexts (RiC) framework \cite{llanes2017records} exemplifies this shift, emphasizing relationships, provenance, and contextual interdependencies across collections rather than isolated descriptions of individual records.

From the perspective of Document Analysis Systems, however, sustaining such a rich and interconnected framework raises practical challenges. The construction and maintenance of contextual knowledge representations (e.g. knowledge graphs) have traditionally relied on the manual labor of experienced historians and archivists. These professionals are expected not only to identify the source and context of a given document, but also to interpret its contents, resolve ambiguities, trace links across collections, and determine the role each record plays within a broader historical narrative. While this work is essential, it is time-consuming, difficult to scale, and inevitably constrained by available resources. This challenge becomes particularly critical in large historical archives containing heterogeneous, partially digitized, and often degraded materials, where access to personalized, well-focused, and detailed information is increasingly the main expectation of researchers and the general public alike. Users no longer seek merely to retrieve documents; they seek explanations, connections, and synthesized insights that cut across fonds, time periods, and institutional boundaries. Meeting these expectations requires systems capable not only of retrieval, but also of modeling and reasoning over knowledge at the collection level. Recent GraphRAG approaches attempt to address multi-hop reasoning and long-term dependencies by integrating graph-based memory structures with language models. However, many of these systems implicitly assume that documents are natively digital, cleanly structured, or at least sufficiently well-processed for LLMs to operate robustly over them. In the context of historical archives, this assumption rarely holds. Documents are frequently handwritten or degraded, transcriptions are noisy, and metadata may be incomplete or inconsistent. As a result, the reliability of downstream reasoning critically depends on the quality and uncertainty management of the transcription and extraction stages.

\begin{figure}[t]
    \centering
    \includegraphics[width=0.9\textwidth]{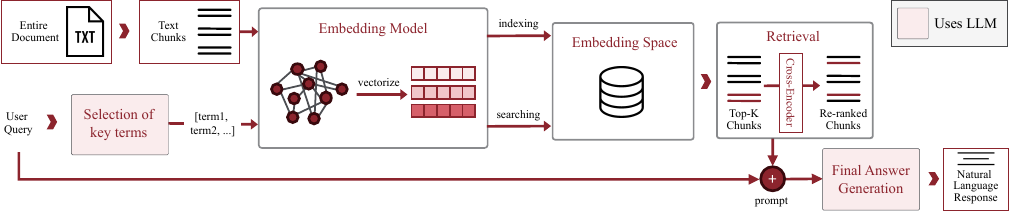}
    \includegraphics[width=0.9\columnwidth]{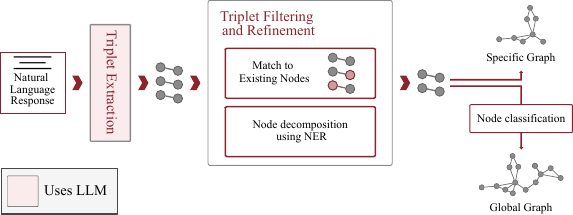}

\caption{Graph-based knowledge modeling framework. Extracted triples build local graphs for conversation context and a global graph linking salient entities across documents.}
    \label{fig:pipeline_retrieval}
\end{figure}

In this work, we address these limitations through two main contributions. First, we introduce a methodology for the on-the-fly creation and continuous refinement of a graph-based knowledge structure during conversational interaction (see Figure~\ref{fig:pipeline_retrieval}). The system stores facts derived either from document retrieval or from expert input into a persistent graph that acts as structured memory for the language model. This graph operates at both a local (conversation-level) and global (collection-level) scale, enabling the accumulation of knowledge across sessions and supporting increasingly complex queries involving multi-document reasoning, link discovery, and contextual interpretation. Second, we propose a transcription enhancement pipeline based on an ensemble of OCR and LLM corrections. Rather than relying on a single automatic transcription, we combine multiple hypotheses and exploit language model completion to reconstruct segments with high uncertainty. By explicitly detecting fragments that can be plausibly reconstructed and flagging those that cannot, the system reduces the risk of propagating hallucinated or low-confidence content into the knowledge layer. 
By combining uncertainty-aware transcription with dynamic graph construction, our approach bridges the gap between document-centered RAG pipelines and knowledge-centered archival frameworks. In doing so, it provides a scalable and interactive pathway toward holistic historical document analysis, aligned both with contemporary advances in language modeling and with long-standing principles in archival science.
For the experimental validation we use a corpus of court-martial sentencing documents from the Spanish Civil War period (1937–1940),

\section{Related Work}

Researchers have long worked on historical document analysis, developing many methods over time. Early efforts focused on Optical Character Recognition (OCR) and Handwritten Text Recognition (HTR), which formed the core of processing pipelines. Still, preprocessing steps like binarization, layout analysis, and text line detection were essential before applying OCR/HTR. The output (plain transcribed text) then required interpretation and information extraction, prompting the use of Named Entity Recognition (NER) and broader Natural Language Processing (NLP) techniques \cite{historical_document_processing}. 

NER \cite{ner_historical} can be considered one of the first and most crucial processing steps for information extraction. It identifies key entities that serve as referential anchors, guiding text interpretation. Yet this approach has its limitations, namely a lack of flexibility: it offers little margin for information retrieval as well as a limited context understanding. The immediate solution was using the NER output as a starting point for Information Extraction (IE) through hand-crafted rule-based and ontology-driven approaches, for tasks such as entity linking and event extraction. With the rise of Transformers, models such as BERT \cite{bert} have been employed to learn semantic patterns, enabling deep contextual understanding and relation extraction. NLP has also greatly benefited from LLMs, especially in retrieval tasks. While models like BERT generate structured outputs, they often lack accessibility. In contrast, LLMs offer a more effective solution for conversational retrieval due to their ability to produce coherent, context-aware responses in natural language. Despite their popularity, LLMs still face challenges, most notably, the issue of 'hallucinations' and a lack of in-domain or specialized knowledge. To address these limitations, Retrieval-Augmented Generation (RAG) was introduced. RAG enhances LLMs by retrieving specific, relevant information to use as contextual input for a query. This allows the LLM \cite{hallucinations_survey} to ground its responses in external knowledge, thereby reducing hallucinations and incorporating domain-specific insights. Advancements in RAG include the use of agents to optimize different components of the system \cite{RAGSurvey}, and the development of hybrid retrieval strategies \cite{Hybrid_RAG} such as sentence-window retrieval and parent-child chunking methods. The emerging challenge lies in effectively retrieving and interpreting complex events defined by who did what to whom, when, and where. Addressing such queries requires a deeper understanding and grounding answers in structured knowledge. To this end, GraphRAG \cite{graph_rag} has been proposed: a fusion of RAG with Knowledge Graphs (KGs), enabling the system to retrieve information from structured sources and present it as contextual input for generation.

The state of the art points toward multi-stage pipelines for historical document processing combined with hybrid RAG/Knowledge Graph systems for information retrieval. This article contributes by integrating these cutting-edge approaches into a single application, providing a novel tool for exploring historically significant archival domains through modern AI techniques.
\section{Method}

\label{sec:dev}


The main objective of our methodology is to transform unstructured and often degraded historical documents into a structured, \textit{queryable} knowledge base that supports both local document analysis and global information synthesis. In contrast to standard RAG architectures, we propose a threefold pipeline which, in conjunction, provides to the system the capacity to accumulate facts within a knowledge graph representation model. 

The implementation of our complete system is organized into three primary components: the text extraction pipeline (Section \ref{sec:transcription}), an advanced RAG-based retrieval system (Section \ref{sec:retrieval}), and a system for the dynamic generation of a Knowledge Graph and its subsequent retrieval (Section \ref{sec:graph}).

\subsection{The Text Extraction Pipeline}
\label{sec:transcription}
\begin{figure}[t]
 \centering
 \includegraphics[width=1\textwidth]{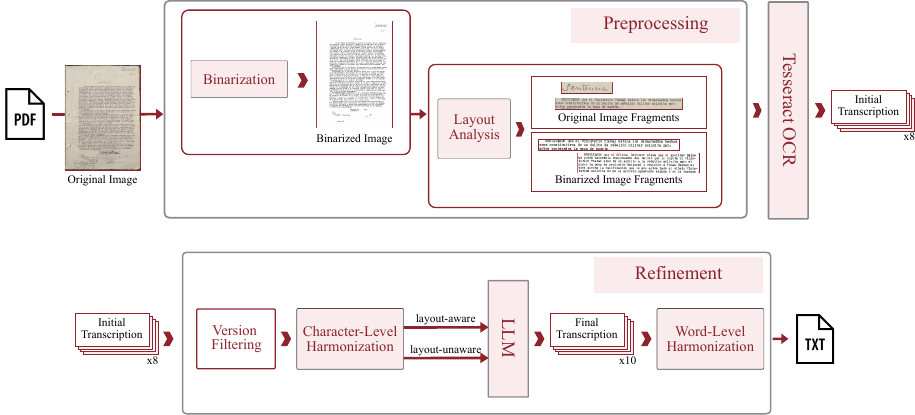}
 \caption{Two-stage transcription pipeline: \textbf{Preprocessing} generates eight OCR candidates via binarization and layout analysis; \textbf{Refinement} filters, harmonizes, and applies LLM-based correction.}
 \label{fig:pipeline_ocr}
\end{figure}

Our multi-stage pipeline, illustrated in Figure \ref{fig:pipeline_ocr}, transcribes historical documents with the highest possible fidelity. It begins with pre-processing to generate multiple OCR candidates, followed by a sophisticated two-step process for consensus and refinement.

Initial analysis of the document corpus highlighted two significant obstacles requiring a robust document analysis pipeline. Firstly, widespread ink-fading and low document contrast frequently caused standard OCR engines to fail, yielding incomplete or empty text. Secondly, the documents exhibit a variable layout, intermittently adopting a two-column format that confounds conventional reading-order algorithms. This structural variability was critical to overcome, as failing to isolate these columns (which contain key subject names) would disrupt the reading order.

To address these challenges, we employ a multi-candidate strategy detailed in Figure \ref{fig:pipeline_ocr}. We generate four base images using distinct binarization techniques (original, global threshold, adaptive threshold, and \textit{Hi-SAM} \cite{HI-SAM}). These inputs feed two parallel processing streams: a \textbf{Direct Path} that extracts text from the raw binaries, and a \textbf{Layout-Aware Path} that utilizes Surya \cite{SURYA} to enforce correct reading order. This yields an ensemble of eight diverse text candidates per document.

This dual-path approach is designed for maximum robustness, with the direct path serving as a crucial failsafe against potential errors in the layout analysis itself. By generating this diverse ensemble of eight candidates, we ensure complete capture of all information from the original text.

Given the complexity of these inputs, the choice of OCR engine was critical. We selected Tesseract after our preliminary analysis showed that it effectively handles the high noise and variable alignment where other solutions failed \cite{historical_document_processing}.




\textit{Final Version Generation}
The raw transcriptions from the OCR stage, while valuable, contain noise and inconsistencies. To produce a single, high-fidelity text, we employ a multi-stage harmonization and refinement process, as shown in the Refinement module of Figure \ref{fig:pipeline_ocr}.

First, the raw OCR outputs are consolidated. We begin by filtering the eight raw transcriptions, discarding outliers based on heuristics such as word count and line length distributions. 

The layout-segmented transcriptions are separated from the others and generated a consensus text within each group using a character-level voting system inspired by \cite{OCR_War}. The system performs a Multiple Sequence Alignment (MSA) to align the texts character by character. At each aligned position, a majority vote determines the final character. This approach is highly effective for correcting common OCR errors, particularly single-character substitutions.

Next, the two baseline transcriptions are passed to an LLM for advanced error correction. Based on extensive evaluation, the `microsoft/phi-4` model \cite{phi4} was selected. To mitigate hallucinations and ensure consistency, each baseline is refined multiple times by the LLM, generating a set of 10 high-quality candidate texts.

Finally, all LLM-refined candidates are reconciled using a word-based voting system (similar to the previous approach but operating on entire words). This step is crucial for preserving the semantic, word-level corrections introduced by the LLM. The final consensus yields a single, highly accurate transcription that best represents the most probable content of the original document.



\subsection{The Advanced RAG}
\label{sec:retrieval}

Once high-fidelity transcriptions are secured, we build a RAG system to answer user queries grounded in the historical documents, following a three-stage design (Figure~\ref{fig:pipeline_retrieval}).

\textit{Offline Indexing.} Document transcriptions are split into semantically coherent chunks following legal section headers like \texttt{RESULTANDO} (resulting, referring to different facts) or \texttt{CONSIDERANDO} (considering, legal considerations or assessments of legal arguments). Then the snippets of text are further segmented into overlapping 50--100 word passages. Each chunk is encoded into a dense vector using \texttt{msmarco-bert-base-dot-v5} from SentenceTransformers~\cite{sentence_transformers}, selected for its strong asymmetric semantic search performance.

\textit{Online Retrieval.} At query time, a \textit{phi-4}-powered agent extracts named entities and reformulates the query into precise search terms. The refined query is encoded and matched against all chunk embeddings via dot-product similarity, retrieving the top-20 candidates. These are re-ranked by a cross-encoder (\texttt{ms-marco-MiniLM-L6-v2}), which jointly scores query--chunk pairs for precise relevance filtering. The top-ranked chunks then vote for a source document, and the most frequently cited document is selected as context.

\textit{Answer Generation.} A prompt combining the original query and retrieved context instructs the model to rely exclusively on the provided evidence, returning a negative response if context is insufficient to prevent hallucinations.
\subsection{Dynamic Knowledge Graph}
\label{sec:graph}

While the RAG system described in Section \ref{sec:retrieval} is effective at retrieving and summarizing information from individual documents, it is limited in its ability to answer complex, relational queries that require synthesizing facts across multiple documents. 
To overcome this limitation, we introduce a dynamic Knowledge Graph (KG) that captures and connects entities and their relationships. The graph is not pre-built but rather dynamically populated based on user interactions. A key novelty of our approach is that the KG acts as a persistent, evolving system memory. Unlike standard RAG systems that process each query in isolation, our graph allows the system to `remember' and accumulate knowledge across multiple user interactions and documents, creating a structured foundation for long-term information synthesis.

\textit{Graph Construction}
\label{sec:graph_construction}
The KG is populated by converting the natural language answers generated by the RAG system into structured (subject, predicate, object) triplets. The core extraction is performed by an LLM agent. We use a dedicated prompt that instructs the model to identify concise, meaningful relationships within the text and return them in a structured list format.

The raw triplets from the LLM are noisy and require significant processing, as illustrated in Figure \ref{fig:pipeline_retrieval}. Predicates that are too long are discarded to ensure relations are meaningful. To prevent duplicate nodes from minor text variations (like capitalization or accents), we find string similarities using normalized Levenshtein distance. New entities are matched against existing nodes, and if a similar enough one is found, the existing node is reused.

Nodes that are overly long or contain conjunctions (like "Juan y Sara") are decomposed. We identify constituent named entities using SpaCy’s NER \cite{spacy} and create a chain of new triplets that explicitly link these components, preserving their original relationships. Before adding the triplets to the graph, we check for duplicates using the string distance method described above, preventing the accumulation of redundant or poorly structured nodes.

This multi-step refinement process transforms raw LLM output into a clean, normalized, and well-structured set of triplets ready for integration into the graph.

\textit{Local and Global Graphs}
A key design choice in our system is the maintenance of two distinct but interconnected knowledge graphs as depicted in the final stage of our creation pipeline (Figure \ref{fig:pipeline_retrieval}).



We maintain two complementary graph structures. A \textbf{local graph} is built per conversation, capturing knowledge extracted from a single document and serving as a structured, queryable representation of its content. A \textbf{global graph} connects all conversations and documents, enabling cross-document synthesis.

\begin{figure}[!t]
    \centering
    \includegraphics[width=0.8\columnwidth]{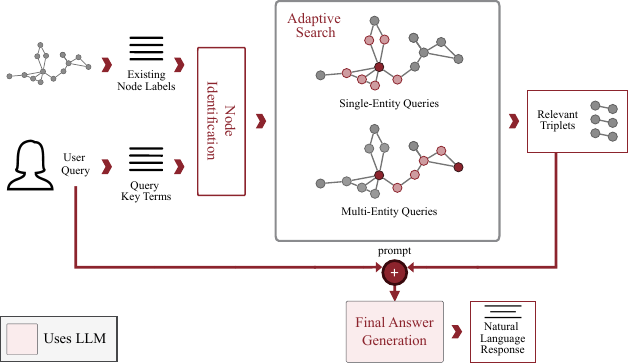}
    \caption{Graph retrieval pipeline: query nodes are identified, relevant triplets retrieved via adaptive search, and used as context for answer generation.}
    \label{fig:pipeline_graph_search}
\end{figure}

The key challenge is merging knowledge across sources without conflating document-specific entities (e.g., the node \textit{``delito''} must represent distinct crimes across documents). To address this, only \textbf{people}, \textbf{places}, and \textbf{dates} --- identified via NER and semantic rules --- are unified globally; all other entities are tagged with a document identifier. This preserves contextual specificity and source traceability while allowing the global graph to link shared entities across documents.
\textit{Graph-Based Retrieval and Question Answering}
When a user asks a question, the system can leverage the Global Graph to find answers that may not be present in any single document. The process for querying the graph and generating a response is detailed in Figure \ref{fig:pipeline_graph_search}.

First, we identify the key aspects of the query (using the same ones identified in the RAG phase) and determine which of these aspects are represented in the Global Graph.

Retrieval then begins by linking the query’s key entities to their corresponding nodes in the graph through string similarity matching, using the same method applied during graph construction.
Based on the number of linked entities identified in the query, the system employs one of two traversal strategies (Figure~\ref{fig:pipeline_graph_search}): for \textbf{single-entity queries} (e.g., \textit{``What is known about Juan Pérez?''}), it retrieves the local neighborhood up to three hops or until a globally relevant entity is found; for \textbf{multi-entity queries} (e.g., \textit{``What is the connection between Juan Pérez and Rafael Prieto?''}), it identifies all simple paths between the corresponding nodes.


To ensure faithful reasoning over the retrieved graph context, we adopt a similar guardrail philosophy as in the RAG component. The LLM is prompted using a graph-specific reasoning template that constrains it to generate answers strictly based on the retrieved triplets. This prompt guides the model to produce a structured JSON response that includes a boolean flag indicating whether a sufficient answer was found, effectively mitigating hallucinations when the graph lacks sufficient evidence.

This graph-based reasoning capability transforms the system from a document retriever into a genuine knowledge exploration tool. To further aid exploration, the system can compute and display global graph analytics, such as node centrality measures, providing users with high-level insights into the entire document collection.

\section{Experimental Setup}

\subsection{Datasets and Tasks}
\label{sec:dataset}
Our work focuses on a corpus of 130 images derived from 65 court-martial sentencing documents from the Spanish Civil War period (1937--1940)~\cite{PARES}. The corpus is of enormous importance for archivists and other heritage professionals because of its significant value in documenting the victims of the civil war and supporting the restoration of collective memory. In this context, it becomes increasingly important to model \textit{who}, \textit{where}, and \textit{why} within the knowledge graph to ensure structured, meaningful, and ethically responsible representation of the past. We use the sentencing pages, typically 1–2 per case, as they summarize key information: details about the accused, individuals involved in the trial, a summary of events, and the final conviction.
\begin{figure}[t]
    \centering
    \begin{subfigure}[t]{0.32\textwidth}
        \centering
        \includegraphics[width=\linewidth]{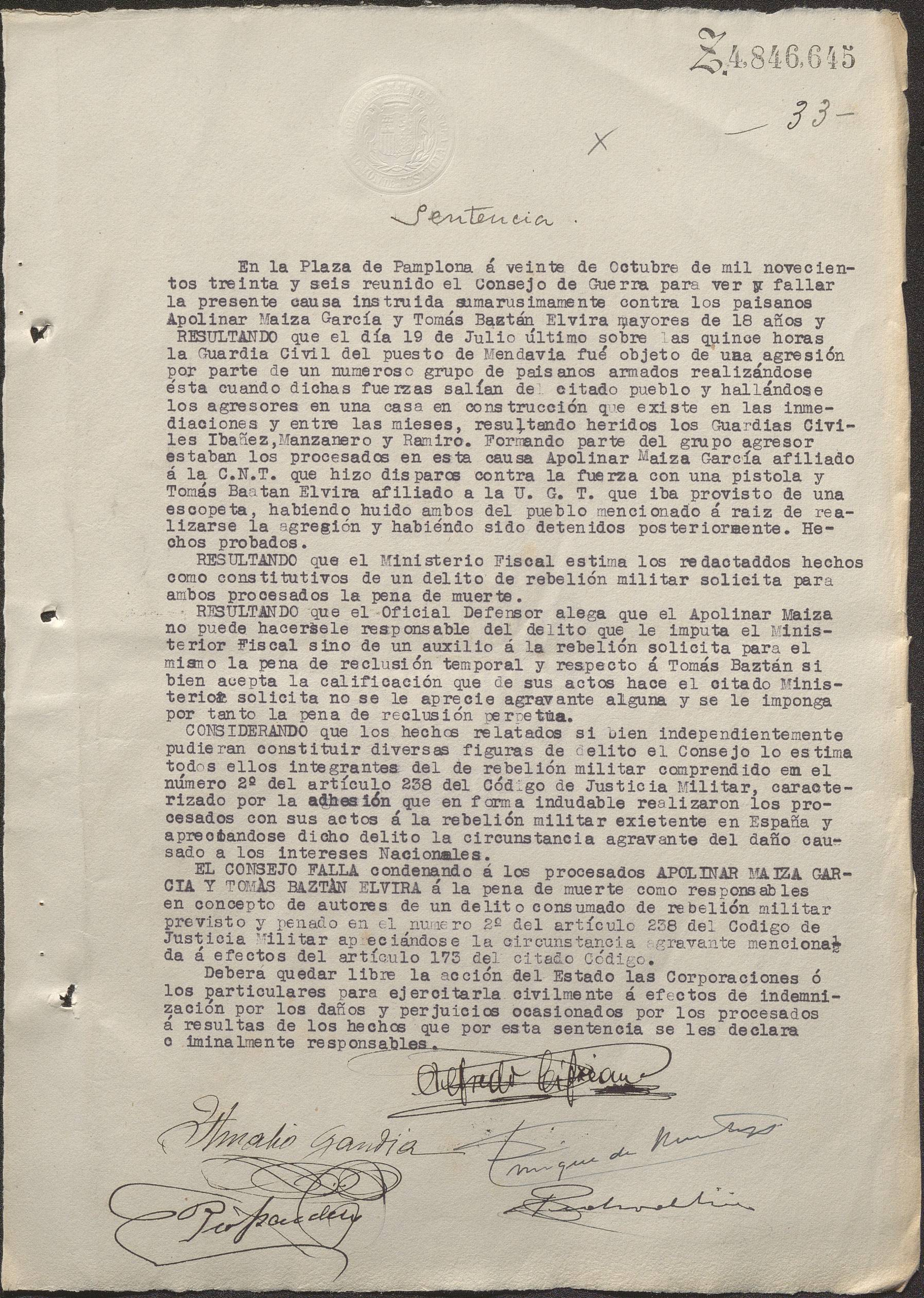}
        \caption{}
        \label{fig:doc-a}
    \end{subfigure}
    \begin{subfigure}[t]{0.32\textwidth}
        \centering
        \includegraphics[width=\linewidth]{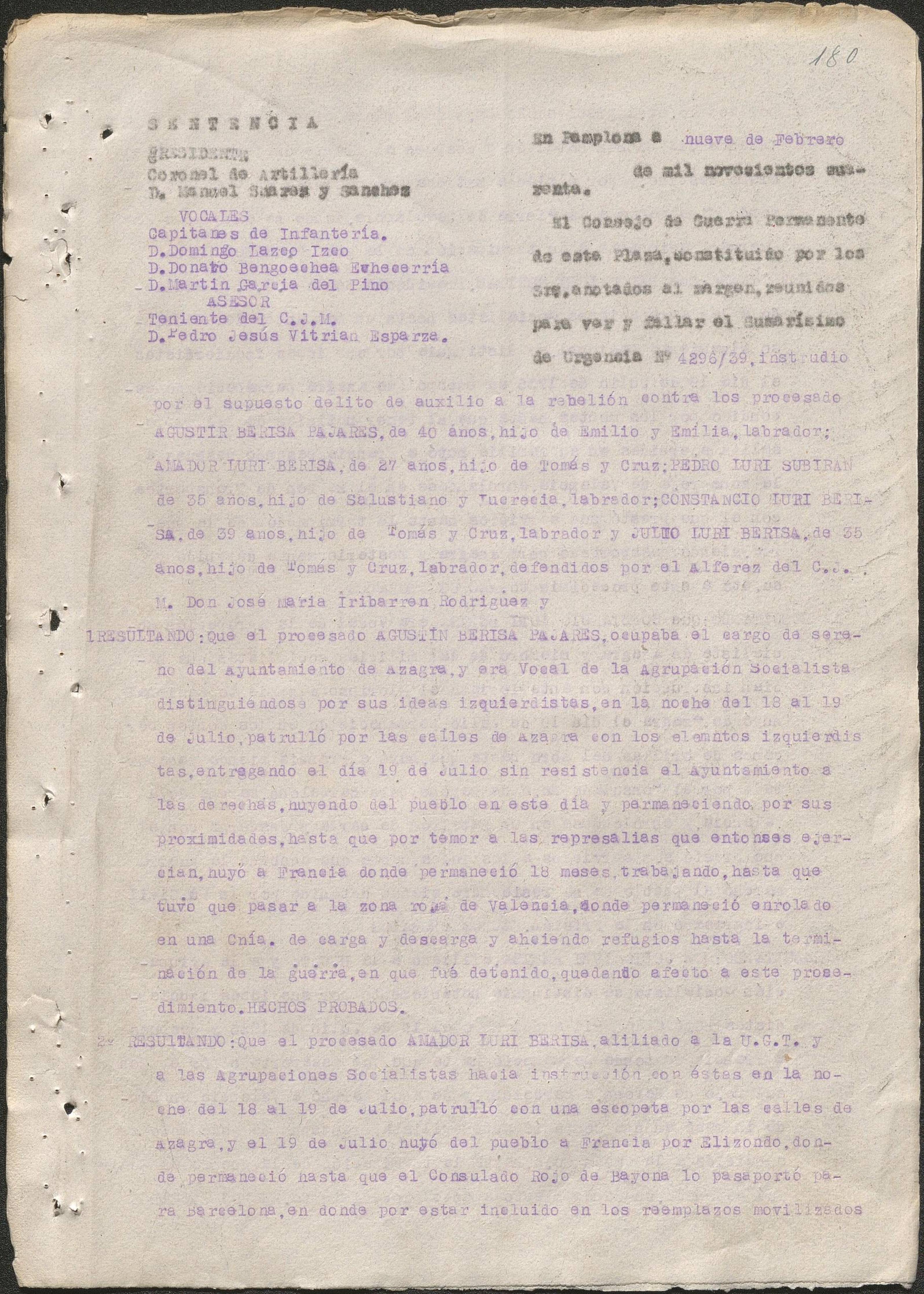}
        \caption{}
        \label{fig:doc-b}
    \end{subfigure}
    \begin{subfigure}[t]{0.32\textwidth}
         \centering
         \includegraphics[width=\linewidth]{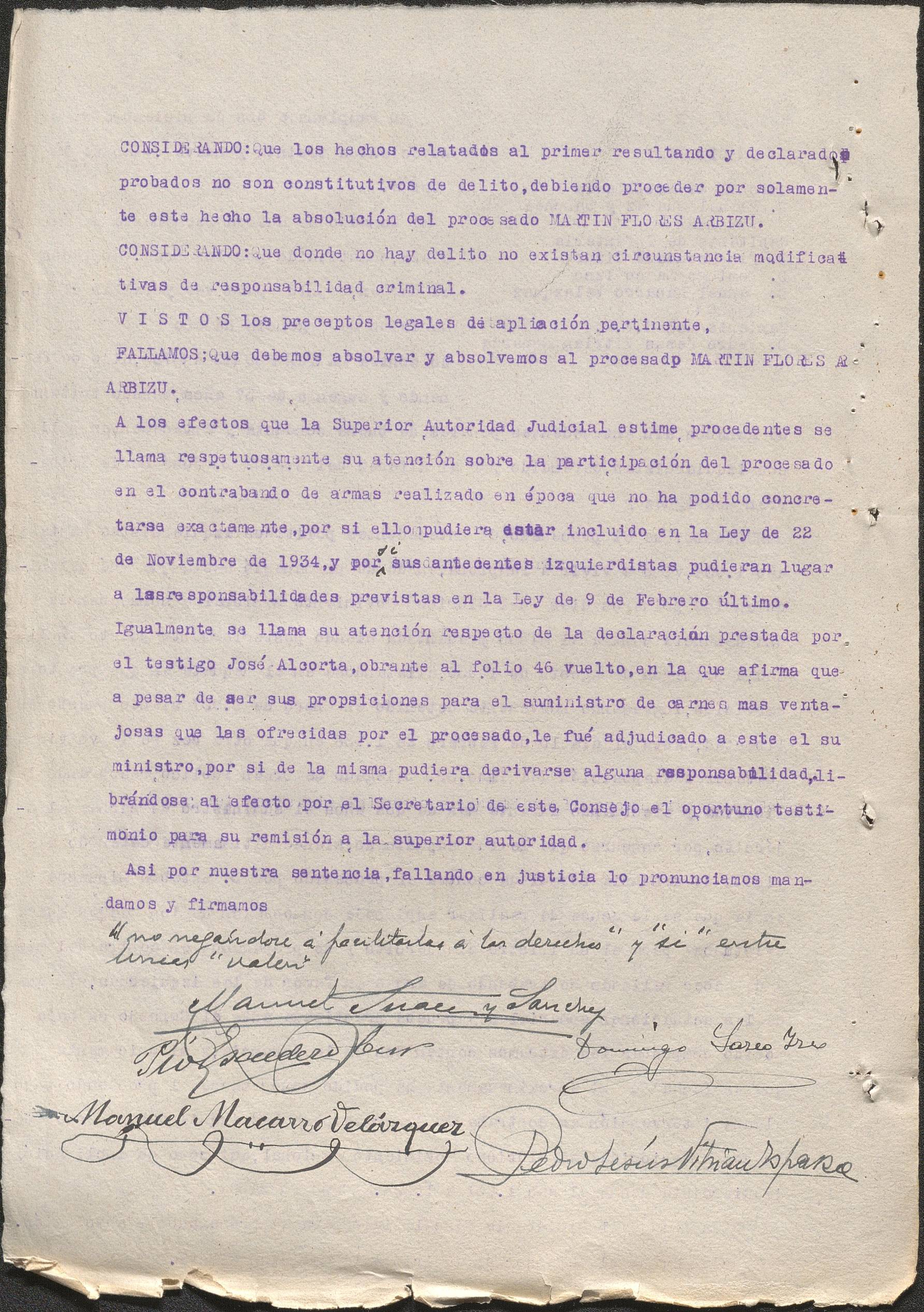}
         \caption{}
         \label{fig:doc-d}
    \end{subfigure}
    \caption{Representative dataset samples: (a) standardized single-column layout with high contrast; (b) complex layout with varying ink, bleed-through, and low contrast; (c) second page of a multi-page document.}
    \label{fig:exemples-documents}
\end{figure}

These historical records present significant text extraction challenges, including faded ink, paper degradation, inconsistent typewriter fonts, and variable layouts with both typewritten and handwritten elements \cite{ImageAnalysis,OCR_War}. As shown in Figure~\ref{fig:exemples-documents}, documents (a-c) illustrate different examples from the working dataset. Document (a) corresponds to a simple sentence with clear text and a straightforward page layout. Even so, handwritten elements, such as the final signatures, are already visible.
Documents (b,c) exemplify the primary technical hurdles of the corpus, specifically multi-page structural complexity and non-uniform background noise. These samples suffer from ink bleed-through (show-through artifacts) and low stroke-to-background contrast (faded ink), which significantly impede binarization and line segmentation algorithms.

\subsection{Metrics and evaluation} 
To validate our system’s performance, we conducted a multi-faceted evaluation. Since there is no existing ground truth dataset for this historical archive, a common challenge in digital humanities, our strategy combines component-wise analysis, qualitative demonstrations, and human-centered studies.

Our evaluation protocol assesses the transcription pipeline first, followed by an automated test of the RAG system, a human evaluation of the retrieval and generation modules, and finally, a visualization of the knowledge graph and conversational interface.

\textit{Evaluation of the Text Extraction Pipeline} We evaluated the multi-stage transcription pipeline using both qualitative and quantitative methods to demonstrate its superiority over a baseline OCR approach.

\textit{Qualitative Analysis} We visually inspected the output to assess how the complete pipeline improves upon the direct application of an OCR engine. This analysis showcases both the significant improvements in layout detection and text fidelity, as well as the limitations that persist in the final output.

\textit{Quantitative Analysis} To quantify the impact of our pipeline, we manually transcribed a representative set of five multi-page documents, including several with severe degradation, to establish a ground-truth (GT) dataset. We compared this generated GT with the output of both the direct OCR application and our complete pipeline. 

We employed Word Error Rate (WER) and Character Error Rate (CER) as our primary metrics. These measure the number of substitutions, insertions, and deletions required to transform the generated text into the ground-truth reference, normalized by the length of the reference. WER operates at the word level, while CER operates at the character level. For both, lower scores indicate better performance.

\textit{Automated Evaluation of Advanced RAG} We conducted an automated evaluation to quantitatively measure the RAG system's performance on fact-retrieval tasks. For this, we used an existing focused ground-truth (GT) dataset containing person-specific attributes, from which we constructed an initial test set of 222 question-answer pairs.

These pairs follow a structured generation pattern based on entity-attribute tuples. For example, given a ground truth entry of \texttt{\{Entity: Juan, Attribute: Age, Value: 22\}}, the corresponding query is ``How old is Juan?''. We will use both the concise extraction (``22'') and natural language variations (``Juan is 22 years old'') as valid targets.

Our evaluation follows a multi-stage funnel to isolate different system components. From the initial 222 queries, our OCR pipeline successfully transcribed the correct source document in 175 cases (a 78.8\% recall). Of these, our RAG agent, which is designed to abstain if context is insufficient, generated a substantive answer for 125 cases. This final set of 125 high-potential responses forms the basis for our in-depth quality and component-level analysis.

To assess accuracy, our primary metric is a \textbf{Custom Exact Match}. This metric employs domain-aware normalization rules to validate if the core factual information was correctly extracted regardless of formatting. For instance, in the age example above, the metric normalizes "twenty-two" to match the ground truth "22", ensuring robust evaluation against linguistic variations.

To further diagnose system behavior, we employed the \textbf{RAGAS framework} \cite{RAGAS,RAG_eval}, which uses an LLM as an automated evaluator across three metrics: \textit{Context Precision} (signal-to-noise ratio of retrieved items), \textit{Context Recall} (coverage of necessary retrieval information), and \textit{Faithfulness} (factual consistency of generated answers against retrieved context).
\textit{Human Evaluation: Evaluating the Efficacy of Dynamic Graphs in Complex Information Synthesis}

Evaluating systems such as RAG or LLMs in question-answering scenarios is challenging without a large ground-truth dataset. For the retrieval component, common metrics include \textbf{relevance} (how well the retrieved context matches the query) and \textbf{accuracy} (how correct the selected documents are among all candidates). For the generation component, metrics typically include \textbf{relevance} (how well the answer addresses the question), \textbf{faithfulness} (how accurately the answer reflects the provided context), and \textbf{correctness} (how close the answer is to an ideal response) \cite{RAG_eval}.
To evaluate our architecture, we conducted a human study with 8 non-expert participants, yielding 64 annotated query-response pairs. We designed two query types: \textbf{Simple Queries (SQ)}, seeking facts about a single entity (e.g., \textit{``What was the sentence for Juan Pérez?''}), and \textbf{Complex Queries (CQ)}, requiring synthesis across multiple documents (e.g., \textit{``Which individuals fled to France?''}).


Participants were presented with answers generated by both the RAG and the Knowledge Graph systems in a blind comparison. First, they were asked to identify whether the RAG system, the Graph system, or both successfully provided an answer. Subsequently, they rated each generated response on a 1–5 Likert scale across four key dimensions: \textbf{Overall Quality}, \textbf{Relevance}, \textbf{Faithfulness}, and \textbf{Answer Depth}.

\subsection{Implementation Details}

\textit{OCR Engine Selection}
To determine the optimal recognition engine for this historical corpus, we qualitatively evaluated seven open-source solutions. Tesseract \cite{tesseract_ocr} performed well on typewritten and binarized documents but struggled with complex layouts. Ocular \cite{ocular_paper}, designed for historical printed or handwritten text using probabilistic models, yielded very poor results on typewritten material and was discarded early. OCRopus \cite{ocropus} worked partially on some images but proved insufficiently robust, failing frequently on uppercase or light text. Calamari \cite{calamari} was similarly discarded after initial testing. Kraken \cite{kraken}, a fork of OCRopus specialized in historical manuscripts with advanced layout analysis, achieved excellent handwritten text recognition but at the cost of slow processing. Ocrad \cite{ocrad}, a lightweight non-neural engine, produced unacceptable results on our documents. Finally, TrOCR \cite{trocr}, a transformer-based encoder-decoder model, showed promise architecturally but performed poorly on this specific historical script without dedicated fine-tuning. Based on this evaluation, Tesseract was selected as the primary engine given its balance of accuracy, robustness to preprocessing, and practical deployability for typewritten court-martial documents.

\begin{figure}[t]
    \centering
    \includegraphics[width=0.9\linewidth]{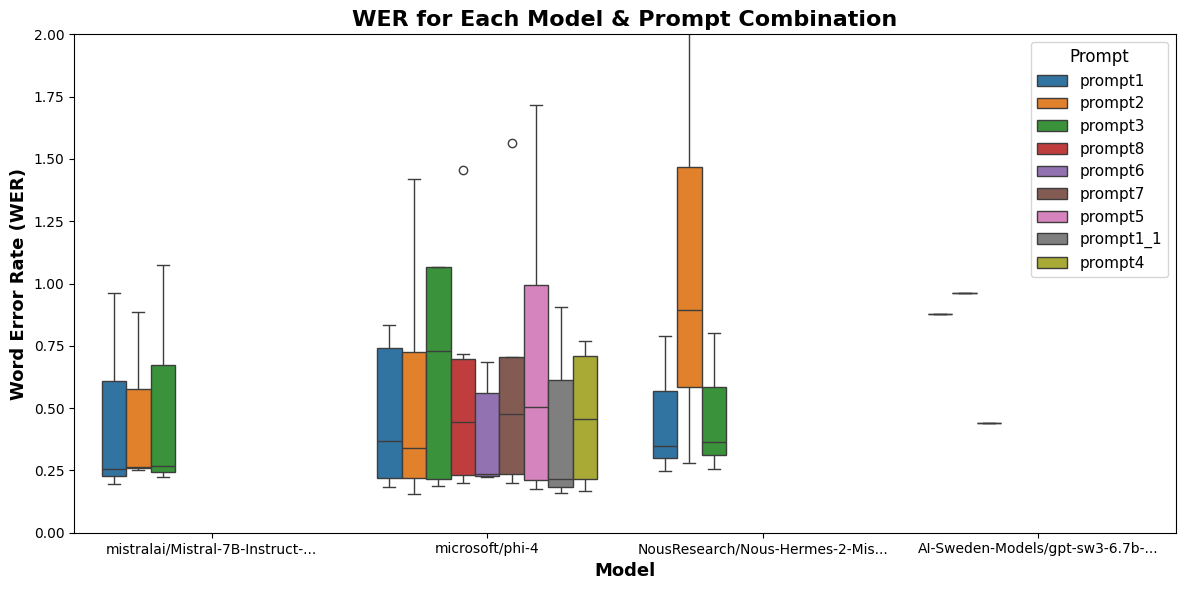} 
    \caption{WER comparison across models and prompts.} 
    \label{fig:grafic_optuna} 
\end{figure}

\textit{LLM Selection and Optimization}
The harmonization and refinement stages rely on a Large Language Model to correct OCR errors without altering semantic meaning. The optimal LLM refinement setup was found via controlled experiments. A small ground truth (GT) dataset was created by manually correcting document transcriptions. Using this GT, various models and prompts were evaluated based on Word Error Rate (WER). Initial tests showed ‘microsoft/phi-4‘ as the best model. Further, prompt engineering optimized its performance. As Figure \ref{fig:grafic_optuna} shows, ``phi-4'' with Prompt 6 achieved the lowest WER and was chosen.

\subsection{Prompt Engineering and Guardrails}

\begin{figure}[t]
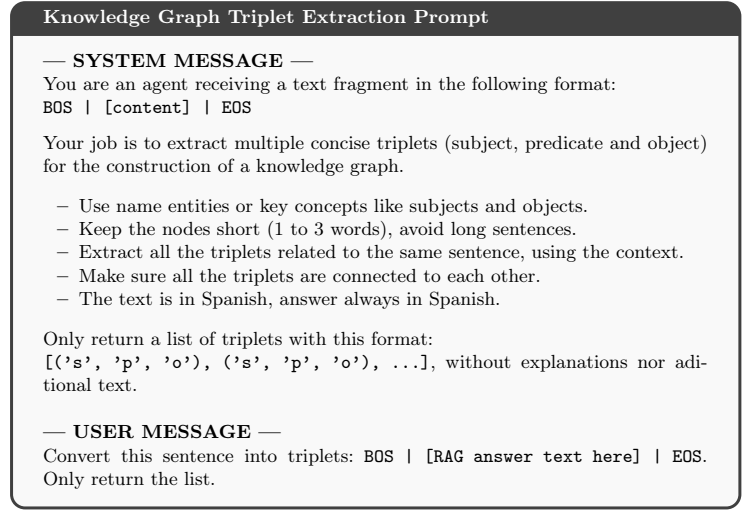

    \centering
    \resizebox{0.8\textwidth}{!}{
    \begin{tcolorbox}[
        colback=gray!5!white,
        colframe=black!75,  
        title=\textbf{Knowledge Graph Triplet Extraction Prompt},
        arc=2mm 
    ]
        \textbf{--- SYSTEM MESSAGE ---} \par
        You are an agent receiving a text fragment in the following format: \\
        \texttt{BOS | [content] | EOS}     
        
        \vspace{0.2cm}
        Your job is to extract multiple concise triplets (subject, predicate and object) for the construction of a knowledge graph.
        \begin{itemize}
            \item Use name entities or key concepts like subjects and objects. 
            \item Keep the nodes short (1 to 3 words), avoid long sentences. 
            \item Extract all the triplets related to the same sentence, using the context. 
            \item Make sure all the triplets are connected to each other.
            \item The text is in Spanish, answer always in Spanish.
        \end{itemize}

        Only return a list of triplets with this format: \\
        \texttt{[('s', 'p', 'o'), ('s', 'p', 'o'), ...]}, without explanations nor aditional text.

        \vspace{0.4cm}
        \textbf{--- USER MESSAGE ---} \par
        Convert this sentence into triplets: \texttt{BOS | [RAG answer text here] | EOS}. Only return the list. 
    \end{tcolorbox}}
    
    \caption{Prompt for extracting \texttt{(subject, predicate, object)} triplets from RAG answers, structured as a \textbf{system message} (rules and output format) and a \textbf{user message} (input text).}
    \label{fig:kg_extraction_prompt} 
\end{figure}

A critical challenge in applying RAG to historical data is preventing the generation of plausible but incorrect facts ("hallucinations"). To address this, we implemented strict \textbf{structural prompting}. Instead of free-text responses, all LLM agents in the system (Query Analysis, RAG Generation, and Graph Extraction) are constrained to output valid JSON. The most critical prompt is the RAG Answer Generator, which includes a mandatory boolean flag (\texttt{found\_information}) forcing the model to explicitly evaluate if the context supports the answer. We also use a prompt to generate the graph triplets (see Figure \ref{fig:kg_extraction_prompt}), which ensures the long sentences are transformed into relevant nodes with significant connections between them. 

\section{Results}
\label{sec:results}

\subsection{Evaluation of the Text Extraction Pipeline}

\begin{figure}[t]
    \centering
    \includegraphics[width=0.95\textwidth]{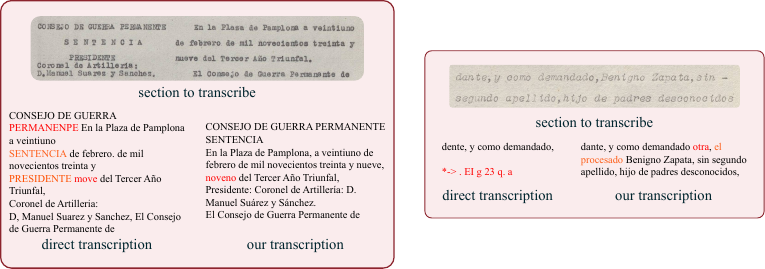} 
    \caption{Two qualitative examples illustrating the improvement from raw OCR output to the final refined text after the complete harmonization and LLM refinement process. Red indicates transcription errors; orange marks layout issues or hallucinated words introduced by the LLM.}
    \label{fig:final_refinement_example}
\end{figure}

\textit{Qualitative Analysis.} Figure \ref{fig:final_refinement_example} shows two qualitative comparisons for representative document snippets. The baseline transcription, produced by applying Tesseract directly to the unprocessed image, contains numerous errors typical of historical documents, such as character substitutions and missed words due to low contrast. In some cases, it completely fails to recognize any text. In contrast, the output from our full pipeline corrects these issues, producing text that is significantly more faithful to the original and layout-aware, although the refinement with the LLM sometimes adds words that do not change the overall meaning but do alter the original transcription.

\textit{Quantitative Analysis.} The difficulty of this corpus is evident in the baseline performance of a standard Tesseract-based OCR, which produces a nearly unusable WER of 72.8\%. In contrast, as shown in Table \ref{tab:transcription_metrics}, our full pipeline reduces the WER to 33.5\%, representing a 54\% relative error reduction. The CER sees a similar improvement, dropping from 46.7\% to 22.4\%. This highlights the critical role of our multi-stage harmonization and LLM-based refinement in achieving high-fidelity transcriptions from difficult historical sources where standard methods fail. However, the WER remains high, partly due to the LLM introducing or substituting words that preserve meaning but alter the original text. It nonetheless indicates that there is still room for improvement.


\begin{table}[b]
\centering
\begin{minipage}[t]{0.32\textwidth}
\centering
\caption{Transcription error rates}
\label{tab:transcription_metrics}
\resizebox{0.975\textwidth}{!}{%
\begin{tabular}{@{}lcc@{}}
\textbf{Method} & \textbf{WER$\downarrow$} & \textbf{CER$\downarrow$} \\
\toprule
Baseline (Tesseract) & 72.8\% & 46.7\% \\
\textbf{Our Pipeline (Full)} & \textbf{33.5\%} & \textbf{22.4\%} \\
\bottomrule
\end{tabular}%
}
\end{minipage}%
\hfill
\begin{minipage}[t]{0.32\textwidth}
\centering
\caption{Evaluation metrics on the filtered set}
\label{tab:evaluation_metrics}
\resizebox{0.975\textwidth}{!}{%
\begin{tabular}{@{}lc@{}}
\textbf{Metric} & \textbf{Score} \\
\toprule
Custom Exact Match & 0.824 \\
\hline
Faithfulness (Generation) & 0.886 \\
Context Precision (Retrieval) & 0.888 \\
Context Recall (Retrieval) & 0.776 \\
\bottomrule
\end{tabular}%
}
\end{minipage}%
\hfill
\begin{minipage}[t]{0.32\textwidth}
\centering
\caption{System answering rate by query type}
\label{tab:activation_rate}
\resizebox{0.975\textwidth}{!}{%
\begin{tabular}{@{}lcc@{}}
\textbf{Query Type} & \textbf{RAG} & \textbf{Graph} \\
\toprule
Simple (N=42) & 83.33\% & 6.25\% \\
Complex (N=14) & 68.75\% & 87.5\% \\
\bottomrule
\end{tabular}%
}
\end{minipage}
\end{table}

\subsection{Automated Evaluation of Advanced RAG}

On the filtered set of 125 questions, we first evaluated end-to-end task success. The Custom Exact metric achieved a high score of 0.824, demonstrating the system's effectiveness. The results of the RAGAS framework provide a more granular view of component performance. The system achieved a very high faithfulness score of 0.886, indicating that the generator is highly reliable and rarely hallucinates. The context precision score of 0.888 further confirms that the model is adept at locating relevant information within the full-document context. Crucially, even within this ``successful'' cohort, the RAGAS context recall metric scored 0.776. This reveals a key insight: while our pipeline identified the correct source document, RAGAS's stricter semantic analysis determined that in 22.4\% of these cases, the context was still insufficient to perfectly formulate the ground-truth answer, often due to subtle OCR errors or the specific information not being present on the document. This highlights the distinction between simple document retrieval and true contextual sufficiency (see Table \ref{tab:evaluation_metrics}).


\subsection{Human Evaluation: Evaluating the Efficacy of Dynamic Graphs in Complex Information Synthesis}

First, we analyzed which system was capable of answering each query type. Table \ref{tab:activation_rate} shows that the systems performed their intended roles. The RAG system handled 83\% of Simple Queries, while the Graph system was nearly silent. Conversely, for Complex Queries, the Graph system activated 87\% of the time, confirming its role as the specialist for relational reasoning. Table~\ref{tab:detailed_combined_scores} presents our core human evaluation results, comparing system performance across both query types. It includes only interactions where the system provided an answer, excluding cases with no response. In addition to the primary quality metrics, we analyzed two specific behaviors of the RAG system based on user feedback: its handling of its own limitations and its robustness to name variations. Table \ref{tab:rag_behavior_analysis} details the frequency of different outcomes for the 51 queries where the RAG system provided an answer.
The results show that the system is generally robust. In over 92\% of cases, it either provided a complete answer or correctly identified and stated its own limitations. Similarly, it successfully handled name variations (or cases where none were needed) over 96\% of the time, with only a small fraction of failures due to entity confusion.

\begin{table}
\centering
\caption{Detailed performance comparison with query types.}
\label{tab:detailed_combined_scores}
\begin{tabular}{@{}ll@{\hspace{8mm}}cccc@{}}
\toprule
 & & \multicolumn{2}{c}{\textbf{Simple Queries (SQ)}} & \multicolumn{2}{c}{\textbf{Complex Queries (CQ)}} \\
\cmidrule(lr){3-4} \cmidrule(l){5-6}
\textbf{Metric} & \textbf{System} & \textbf{Mean} & \textbf{Std. Dev.} & \textbf{Mean} & \textbf{Std. Dev.} \\
\midrule
Overall Quality & RAG & \textbf{4.50} & 1.18 & 4.64 & 0.81 \\
 & Graph & 3.67 & 2.31 & \textbf{4.71} & 0.83 \\
\midrule
Relevance & RAG & \textbf{4.67} & 0.99 & 3.73 & 1.42 \\
 & Graph & 3.67 & 1.53 & \textbf{4.86} & 0.53 \\
\midrule
Faithfulness & RAG & 4.82 & 0.68 & 4.36 & 1.43 \\
 & Graph & \textbf{5.00} & 0.00 & \textbf{4.64} & 1.08 \\
\midrule
Answer Depth & RAG & 4.61 & 0.82 & 4.45 & 0.82 \\
 & Graph & \textbf{4.67} & 0.58 & \textbf{4.64} & 0.84 \\
\bottomrule
\end{tabular}
\end{table}

For Simple Queries, the RAG system significantly outperforms the Graph in both Overall Quality and Relevance. While scores for Faithfulness and Answer Depth are comparable, RAG's superior relevance confirms its role as the effective baseline for single-document questions. For Complex Queries, the results highlight the critical role of the Knowledge Graph. The most striking result is the reversal in the Relevance metric, where the Graph system is clearly superior.Furthermore, the Graph demonstrates stronger performance in Faithfulness and Answer Depth. These results provide strong evidence that while RAG can generate fluent answers, the Graph is essential for retrieving the correct cross-document context needed for genuine synthesis, confirming the complementary strengths of our hybrid architecture. In summary, the human evaluation confirms the specialized roles of our architecture. Its overall effectiveness is reflected in the final user satisfaction score, which averaged 4.12 out of 5, indicating a highly positive user experience even when accounting for system failures.
\begin{table}[t]
\centering
\caption{Detailed Behavior Analysis of the RAG System }
\label{tab:rag_behavior_analysis}
\begin{tabular}{@{}llc@{}}
\textbf{Category} & \textbf{Metric (Observed Outcome)} & \textbf{Frequency} \\
\toprule
\textbf{Limitations Handling} & Complete (No limitations apply) & 0.745 \\
& Stated limitations clearly & 0.177 \\
& Failed to state limitations & 0.078 \\
\hline
\textbf{Name Variation Handling} & Handled correctly (No variation) & 0.922 \\
& Handled correctly (Variation present) & 0.039 \\
& Failed (Entity confusion) & 0.039 \\
\bottomrule
\end{tabular}
\end{table}



\section{Conclusions}
We have validated a system for automatic knowledge modeling via conversational interfaces, comparing our graph-based approach against standard RAG pipelines. Among seven OCR engines evaluated, \textit{Tesseract} offered the best trade-off between efficiency and performance. An ensemble-based correction methodology reduces hallucinations and improves robustness in downstream knowledge extraction. Evaluation shows that GraphRAG and standard RAG perform comparably on simple extractive queries, but GraphRAG demonstrates clear superiority on complex, multi-hop reasoning across all metrics. Beyond quantitative gains, the key contribution is enabling on-the-fly knowledge modeling in historical digital libraries: expert archivists can iteratively refine domain knowledge through conversation, transforming the interface from a retrieval tool into an evolving knowledge management environment. By unifying primary sources with curated and inferred facts in a structured graph index, the system supports long-term dependencies, cross-record linking, and contextual expertise not explicitly encoded in documents. Future work will address scalability, more sophisticated graph reasoning, and deeper human-model collaboration workflows.

\section*{Acknowledgements}
\footnotesize{
This work has been partially supported by the Spanish project PID2024-157778OB-I00, Ministerio de Ciencia e Innovación, the Departament de Cultura of the Generalitat de Catalunya, and the CERCA Program. Adrià Molina is funded with the PRE2022-101575 grant provided by MCIN / AEI / 10.13039 / 501100011033 and by ERDF/EU.
This work was partially included in Paula's Bachelor's Thesis and can be found in~\cite{tfg}.}

\bibliographystyle{splncs04}
\bibliography{mybibliography}

\end{document}